\documentstyle[psfig]{mn}
\newif\ifAMStwofonts
\def\kms{$\mathrm kms^{-1}$}
\def\ll{$\lambda$}
\def\zabs{$z_{abs}$}
\chardef\ii="10
\ifoldfss
  \ifCUPmtlplainloaded \else
    \NewTextAlphabet{textbfit} {cmbxti10} {}
    \NewTextAlphabet{textbfss} {cmssbx10} {}
    \NewMathAlphabet{mathbfit} {cmbxti10} {} % for math mode
    \NewMathAlphabet{mathbfss} {cmssbx10} {} %  "   "    "
  \fi
  \ifAMStwofonts
    \ifCUPmtlplainloaded \else
      \NewSymbolFont{upmath} {eurm10}
      \NewSymbolFont{AMSa} {msam10}
      \NewMathSymbol{\upi}     {0}{upmath}{19}
      \NewMathSymbol{\umu}     {0}{upmath}{16}
      \NewMathSymbol{\upartial}{0}{upmath}{40}
      \NewMathSymbol{\leqslant}{3}{AMSa}{36}
      \NewMathSymbol{\geqslant}{3}{AMSa}{3E}

    \fi
  \fi
\fi % End of OFSS

\ifnfssone
  \newmathalphabet{\mathit}
  \addtoversion{normal}{\mathit}{cmr}{m}{it}
  \addtoversion{bold}{\mathit}{cmr}{bx}{it}
  \newmathalphabet{\mathbfit} % math mode version of \textbfit{..}
  \addtoversion{normal}{\mathbfit}{cmr}{bx}{it}
  \addtoversion{bold}{\mathbfit}{cmr}{bx}{it}
  \newmathalphabet{\mathbfss} % math mode version of \textbfss{..}
  \addtoversion{normal}{\mathbfss}{cmss}{bx}{n}
  \addtoversion{bold}{\mathbfss}{cmss}{bx}{n}
  \ifAMStwofonts
    \ifCUPmtlplainloaded \else
      \UseAMStwoboldmath
      \makeatletter
      \new@mathgroup\upmath@group
      \define@mathgroup\mv@normal\upmath@group{eur}{m}{n}
      \define@mathgroup\mv@bold\upmath@group{eur}{b}{n}
      \edef\UPM{\hexnumber\upmath@group}
      \new@mathgroup\amsa@group
      \define@mathgroup\mv@normal\amsa@group{msa}{m}{n}
      \define@mathgroup\mv@bold\amsa@group{msa}{m}{n}
      \edef\AMSa{\hexnumber\amsa@group}
      \makeatother
      \mathchardef\upi="0\UPM19
      \mathchardef\umu="0\UPM16
      \mathchardef\upartial="0\UPM40
      \mathchardef\leqslant="3\AMSa36
      \mathchardef\geqslant="3\AMSa3E
    \fi
  \fi
\fi % End of NFSS release 1

\ifnfsstwo
  \DeclareMathAlphabet{\mathbfit}{OT1}{cmr}{bx}{it}
  \SetMathAlphabet\mathbfit{bold}{OT1}{cmr}{bx}{it}
  \DeclareMathAlphabet{\mathbfss}{OT1}{cmss}{bx}{n}

  \SetMathAlphabet\mathbfss{bold}{OT1}{cmss}{bx}{n}
  \ifAMStwofonts
    \ifCUPmtlplainloaded \else
      \DeclareSymbolFont{UPM}{U}{eur}{m}{n}
      \SetSymbolFont{UPM}{bold}{U}{eur}{b}{n}
      \DeclareSymbolFont{AMSa}{U}{msa}{m}{n}
      \DeclareMathSymbol{\upi}{0}{UPM}{"19}
      \DeclareMathSymbol{\umu}{0}{UPM}{"16}
      \DeclareMathSymbol{\upartial}{0}{UPM}{"40}
      \DeclareMathSymbol{\leqslant}{3}{AMSa}{"36}
      \DeclareMathSymbol{\geqslant}{3}{AMSa}{"3E}
    \fi
  \fi
\fi % End of NFSS release 2

\ifCUPmtlplainloaded \else
  \ifAMStwofonts \else % If no AMS fonts
    \def\upi{\pi}
    \def\umu{\mu}
    \def\upartial{\partial}
  \fi
\fi

\title[Oscillator strength of S{\sc ii} line]{   
An astrophysical  oscillator strength for the  S{\sc ii}
94.7 nm\\ resonance line and S abundances in DLAs .
\thanks{based on data from the ORFEUS II mission,
the Hubble Space Telescope, the International
Ultraviolet Explorer  and the ESO Very Large Telescope}
}
\begin{document}
\author[Bonifacio et al ]{Piercarlo Bonifacio$^1$,
Elisabetta Caffau$^2$, 
Miriam Centuri\'on $^1$,
Paolo Molaro $^1$,
\newauthor
Giovanni Vladilo $^1$
\\
$^1$Osservatorio Astronomico di Trieste,Via G.B.Tiepolo 11,
I-34131 Trieste, Italy
\\
$^2$Istituto Magistrale S.P.P. e L. annesso al Convitto Nazionale
Paolo Diacono, S. Pietro al Natisone
Udine, Italia
}
%\offprints{P. Bonifacio}
\date{received .../Accepted...}
\maketitle

\label{firstpage}
\begin{abstract}

By using   UV spectra for
the O star HD 93521 taken with the ORFEUS II echelle spectrograph,
we determine an ``astrophysical'' $f$ value
for
the S{\sc ii} \ll 94.7 nm line:
$f  = 0.00498 {}^{+0.00172}_{-0.00138} $, error at 1 $\sigma$
level. 
 This is almost a factor of 30 smaller than 
the guessed value found in the Kurucz
database ($f=0.1472$), which  
was up to now the  only one available for this
transition.
We use our
``astrophysical'' $f$  to investigate  the S abundance in 
two Damped Ly$\alpha$ absorption systems (DLAs)
observed with UVES at the ESO 8.2m Kueyen telescope.
In the case of 
the absorber at $z_{abs}=3.02486$
towards QSO 0347-3819 we find a sulphur column density
which is consistent, within errors, with that determined by Centuri\'on et al
by means of the \ll 125.9 nm line, thus providing an external check
on the accuracy of our $f$ value.
For the damped
absorber at 
$z_{abs}=4.4680$ towards BR J0307-4945
we  determine a high value of 
the S abundance,  which, however, 
is probably the result of blending with Ly$\alpha$ 
forest lines.

\end{abstract}

 
\section{Introduction}

In the chemical
study of DLAs the importance of S is 
twofold: first it traces the $\alpha$
elements, which are synthesized mainly by Type II SNe, so that  the
$\alpha$/iron--peak ratio provides an important constraint
on the chemical history of the observed material, and second 
S is known to be essentially undepleted onto dust grains
(Savage \& Sembach 1996),
at variance with what happens for Si, which is
the $\alpha$ element
most  commonly observed in DLAs. 
Therefore the S abundance in the gas phase
measures the actual S abundance in the system
(Centuri\'on et al 2000).
In the spectrum of  BR J0307-4945,
obtained with UVES on the ESO-Kueyen 8.2 m telescope
during the commissioning of the instrument,
a DLA at \zabs = 4.4680 is observed.
For this system the  S{\sc ii} triplet lines 
at 125 nm, which are the only ones used so  far to determine
the abundances of this nucleosynthetically important element
(Centuri\'on et al 2000), are not available because they 
are blended with lower redshift Ly$\alpha$ absorbers.
We have been able to 
detect the  S{\sc ii} \ll 94.7 nm
resonance line for which,
unfortunately, no theoretical or experimental $f$ value
is available, thus preventing its use in the  analysis.
The presence of the  S{\sc ii} \ll 94.7 nm line 
and the lack of a suitable $f$
value are tantalizing. 

The observations
with the echelle spectrograph on board of the ORFEUS II mission
allow to observe the  S{\sc ii} \ll 94.7 nm line
in the Galactic interstellar medium
with both a resolution and signal to noise ratio sufficient for its
analysis. We therefore decided to use these data
to derive an ``astrophysical''
$f$ value for this transition.
Our method is quite simple: 
we determine the $f$ value by requiring that
this line yields the same column 
density as the  S{\sc ii} \ll 125 nm triplet. 
The most suitable target for this analysis is the well known O star
HD 93521.  The  S{\sc ii} \ll 94.7 nm line has
been clearly detected in its ORFEUS II echelle 
spectrum by Barnstedt et al (2000).
In the present paper we perform a quantitative analysis on this feature.
The ORFEUS spectrum covers both the  S{\sc ii} \ll 94.7 nm line and the
S{\sc ii} \ll 125 nm triplet. In order
to support our analysis we also make use of archival data from
IUE and HST-GHRS both of which contain 
information on the  S{\sc ii} \ll 125 nm triplet.

\section{Observational material}

Our data consist of two echelle spectra from the ORFEUS II
mission, three echelle spectra observed  with the GHRS on board 
the HST.
We also used  128 high resolution large aperture SWP IUE spectra. 

\begin{figure}
\psfig{figure=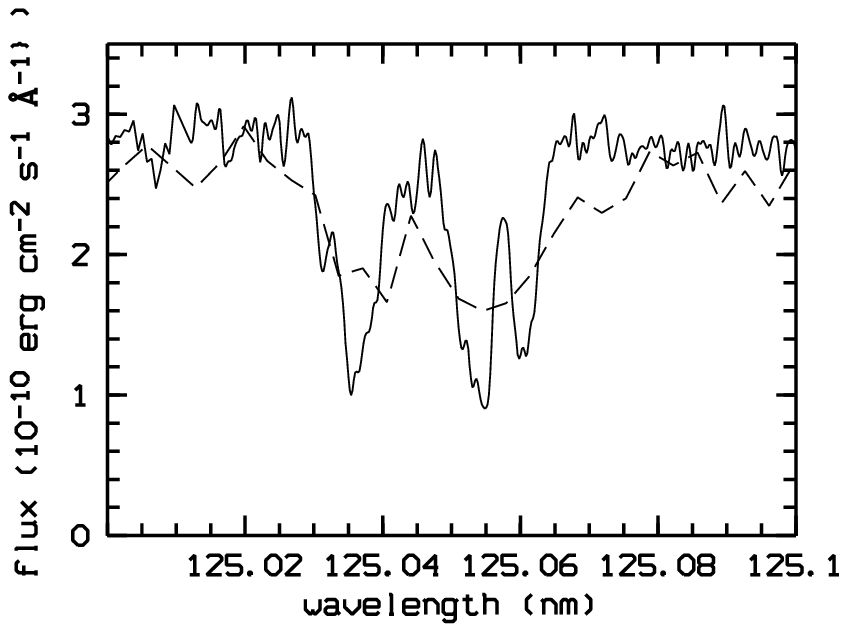,clip=t}
\caption{The  S{\sc ii} \ll 125.0 nm line with HST (solid line)
and ORFEUS (dashed line).}
\label{s1250}
\end{figure}

\begin{figure}
\psfig{figure=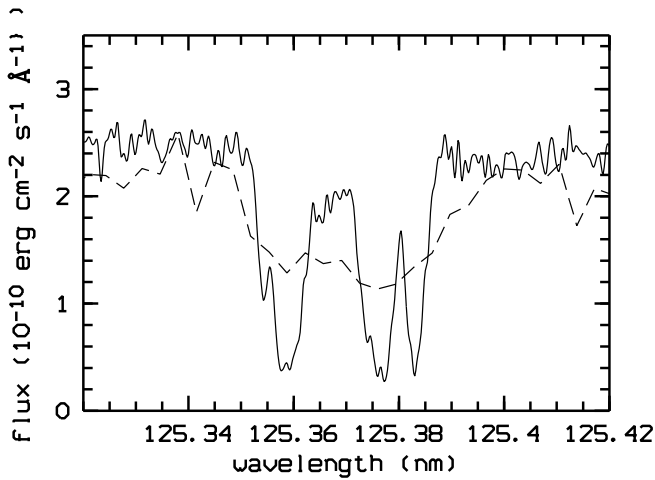,clip=t}
\caption{The S{\sc ii}\ll 125.4 nm line with
HST (solid line) and ORFEUS (dashed line).}
\label{s1253}
\end{figure}

The ORFEUS data under study here
were described in detail by Barnstedt et al
(2000) and instrument description, performance and data
reduction by Barnstedt et al (1999). 
The wavelength coverage ranges from about 91.6 nm to 141.0 nm
with a resolution of $\lambda /\Delta\lambda \approx 10000$.
Each ORFEUS echelle spectrum consists of 22 echelle orders.
We focussed on order 44 (125.421 nm -- 128.457 nm),
order 45 (122.635 nm -- 125.602 nm) and order 59
(93.500 nm -- 95.999 nm).
Orders 44 and 45 contain the  S{\sc ii} \ll 125 nm triplet
while order 59 contains the  S{\sc ii} \ll 94.7 nm line.
The ORFEUS data was kindly provided to us by
Dr. Kappelmann. 

HST data consist of three GHRS echelle spectra, taken
with the small science aperture
(z0ih020ft, z0ih020gt, z0ih020ht), which we added without any wavelength 
shift.
For each spectrum
the resolution is $\lambda /\Delta\lambda \approx 85000$
and the exposure time is 144 s.
These spectra were already analyzed by Spitzer \& Fitzpatrick (1993).

\begin{figure}
\psfig{figure=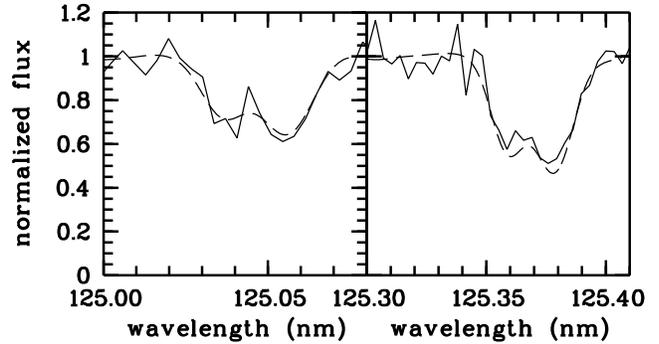,clip=t}
\caption{Normalized flux of ORFEUS (solid line) and
HST (dashed line) of two of the S{\sc ii} \ll 125 nm  triplet lines.
The  HST data have been broadened to the resolution of ORFEUS
and shifted in wavelength to match the  125.4 nm line
which lie in order 45.
}
\label{norm}
\end{figure}

The UVES spectra of the QSOs 
BR J0307-4945
and
0347-3819 
were obtained during the commissioning of the instrument and
released by ESO for public use. 
In the present paper we use one spectrum 
of BR J0307-4945 taken in the red arm centered at 600 nm with
CD \#3 and a 1\farcs{1} slit and an exposure time of 4400 s.
For QSO 0347-3819 we use
the sum of two blue arm spectra, for a total
exposure time of 9500 s taken with dichroic \#2 and centered at 437 nm
CD \#2 and a 1\farcs{0} slit.
The reduced spectra were provided to us by 
Dr. M. Dessauges-Zavadsky and Dr. S. D'Odorico  
who used the UVES data reduction pipeline.
Further details  on the  data reduction may be found
in Dessauges-Zavadsky et al (2001) and D'Odorico et al (2001).
The resolution is about 40000 for both spectra 
with S/N $\sim 40 $ in the regions under analysis here.

\section{Analysis and results}

Since
HD 93521 is a fast rotator ($v \sin i = 400 \pm 25 $ \kms,
Lennon et al 1991) 
all stellar lines are very broad
and it is therefore relatively easy to draw a local continuum
relative to which the interstellar line may be measured. 
When we refer to ``lines'' in the following discussion we always mean the
interstellar lines.

Due to  the differences in resolution
we could not compare  HST data with ORFEUS spectra directly.
Even the wavelength scale of the two instruments is not in agreement,
as may be appreciated in figures \ref{s1250}--\ref{s1253}.
By least squares fitting we determined the wavelength shift and broadening
to be applied to  HST spectra in order to obtain
the best agreement with the ORFEUS spectrum.
In Fig.\ref{norm} the degraded HST  normalized spectrum 
is compared  to the corresponding ORFEUS spectrum.
The spectra have been normalized by fitting a spline through
selected continuum points.

Although the HST full resolution spectra show the presence of 9
separate components (Spitzer \& Fitzpatrick 1993), 
at the
resolution of ORFEUS only  two components are apparent,
indicated hereafter as B (blue) and R (red) component.
We fitted Voigt profiles to the spectra of the two instruments by
using least squares
in order to obtain the column density, the $b$ value
and the mean velocity
of the two components of each line of the  S{\sc ii} \ll 125 nm
triplet.  
The atomic data for the transitions under study are summarized in
Table \ref{atomdat}.
The results of the fits are given in Table \ref{par}.

A similar exercise was performed also using IUE data, 
for which we coadded 128 IUE
large aperture SWP spectra. The flux of these spectra appeared 
in disagreement with that of both ORFEUS and HST. We believe this 
is due to the well known problems of background subtraction in IUE at 
these short wavelengths. We thus did not further consider the IUE spectrum.

\begin{table}
\caption{Atomic data}
\label{atomdat}
\begin{center}
\begin{tabular}{rrrrr}
\hline

 $\lambda _{\circ}$ & $f$ & $g_1$ & $g_2$ & ref \\
 nm  &  \\
\hline
 94.6978 & 0.00498 & 4.0 & 4.0 & this paper \\
125.0584 & 0.00545 & 4.0 & 2.0 & Morton 1991\\
125.3811 & 0.01088 & 4.0 & 4.0 & Morton 1991\\
125.9519 & 0.01624 & 4.0 & 6.0 & Morton 1991\\
\hline
\end{tabular}
\end{center}
\end{table}

\begin{table}
\caption{Best fit parameters for the S{\sc ii} 125 nm triplet}
\label{par}
\begin{center}
\renewcommand{\tabcolsep}{3pt}
\begin{tabular}{rrrrrr}
\hline
line & N & b      & N & b      & $\Delta v$ \\
    & \multispan2{~~Blue~}& \multispan2{~~~Red} & 
{($\mathrm kms^{-1}$)}\\
\hline
\multispan2{ORFEUS}\\
\\
125.0 & $14.91$ & $10.52$ & $15.02$& $13.05$ & -45.52\\
 & $\pm 0.09$ & $\pm 2.5$ & $\pm 0.07$ & $\pm 3.0$ & $\pm 1.0$\\
125.4 & $14.91$ & $6.10$ & $15.02$ & $17.03$ & -45.52\\
 & $\pm 0.09$ & $\pm 2.0$ & $\pm 0.07$ & $\pm 7.0$ & $\pm 1.0$ \\
125.9 & $14.91$ & $10.52$ & $15.02$ & $13.05$ & -47.22\\
 & $\pm 0.09$ & $\pm 2.5$ & $\pm 0.07$ & $\pm 3.0$ &  $\pm 1.0$\\
\hline
\multispan2{HST}\\
\\
125. & $ 14.86 $ & $12.14$ & $ 15.03 $ & $17.27$ & -45.84\\
     & $ \pm 0.01 $ & $\pm 1.0$ & $ \pm 0.01 $ & $\pm 1.4$ & $\pm 0.1$\\
\hline
\end{tabular}
\end{center}
\end{table}

\begin{figure}
\psfig{figure=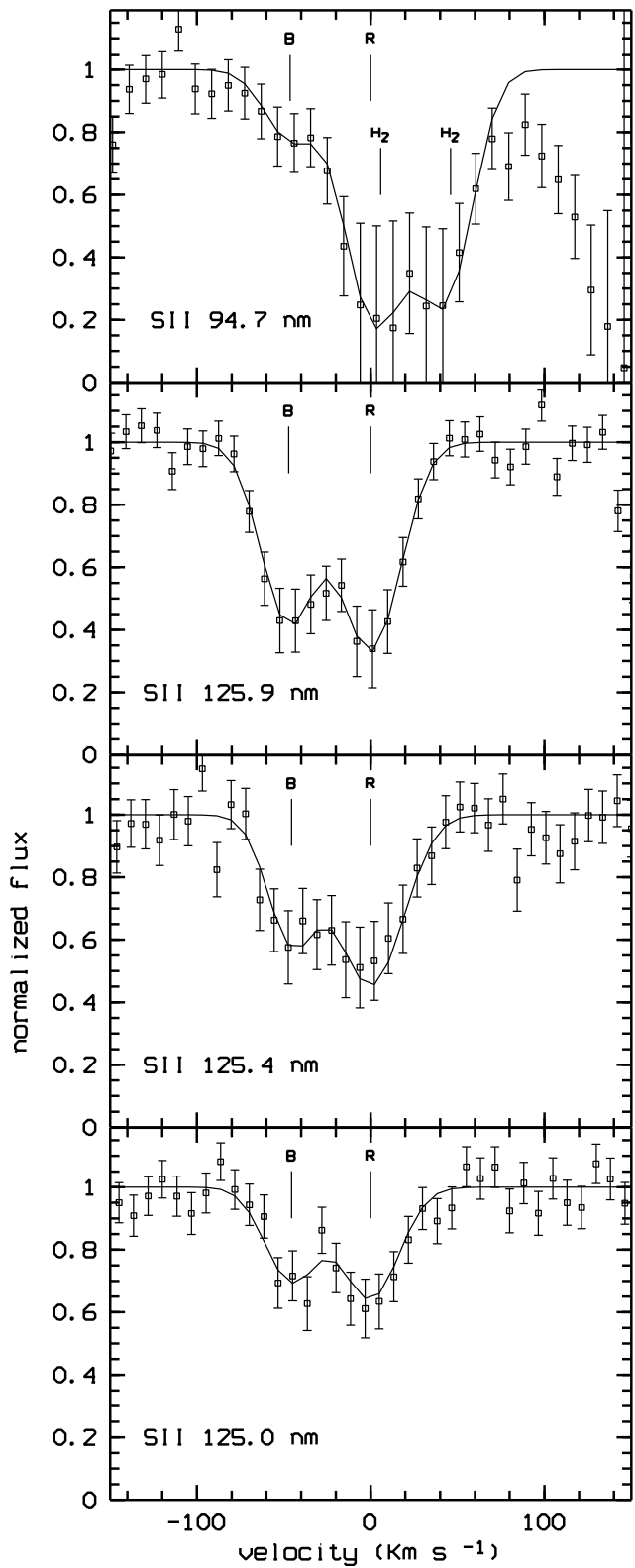,clip=t}
\caption{The synthetic spectra of the S{\sc ii} lines
are over imposed
to the data (squares with error bars). 
B and R identify the center of the blue ($v=-47$ \kms) and red
($v=0$ \kms) components. In the top panel also the two H$_2$
blending lines are indicated.}
\label{fit}
\end{figure}

Using the column density, the $ b$ parameter and the mean velocity 
for the B and R component of the best fit to the triplet lines
we fitted 
the whole 
range from 94.6680 nm to    94.7120 nm
in order to determine the $f$ value of the S {\sc ii}
\ll 94.7 nm line.

We performed the fitting starting
with the input parameters deduced from
ORFEUS and HST in turn. 
The R component of the S{\sc ii} 94.7 nm  line
is blended with the $\rm H_2$ \ll 94.6986 nm line.
For the $\rm H_2$ lines we fixed $b=7{\mathrm km s^{-1}}$,
as found for all other  $\rm H_2$ lines by Gringel
et al (2000), while we left the column densities as fitting parameters.
For what pertains to the S{\sc ii} \ll 94.7 nm
lines all the parameters where kept fixed
except  the $f$ value.
Therefore we had three fitting parameters and 12 degrees of freedom;
the results are reported in Table \ref{par}.

Taking as input the column densities
derived from the ORFEUS data the fitted oscillator strength 
of the S {\sc ii} \ll 94.7 nm line is 

$$f = 0.00498 {}^{+0.00172}_{-0.00138} $$

\begin{figure}
\psfig{figure=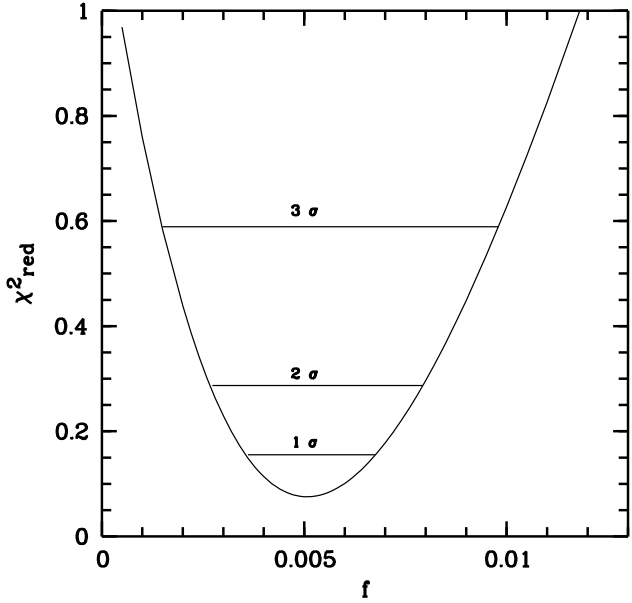,clip=t}
\caption{The reduced $\chi ^2$ as a function of $f$.}
\label{chiq}
\end{figure}

The fit is shown in Fig. \ref{fit}.
The quoted error is the 1$\sigma $ 
deduced from the $\chi^2$ analysis, as shown
in Fig.\ref{chiq}.
The oscillator strength found by
using the column density deduced from HST data is 
$f = 0.00528{}^{+0.00157}_{-0.00148}$, in good 
agreement
with that 
obtained above by
using the  ORFEUS--based column density.

\section{Application to the study
of QSO 0347-3819}

\begin{figure}
\psfig{figure=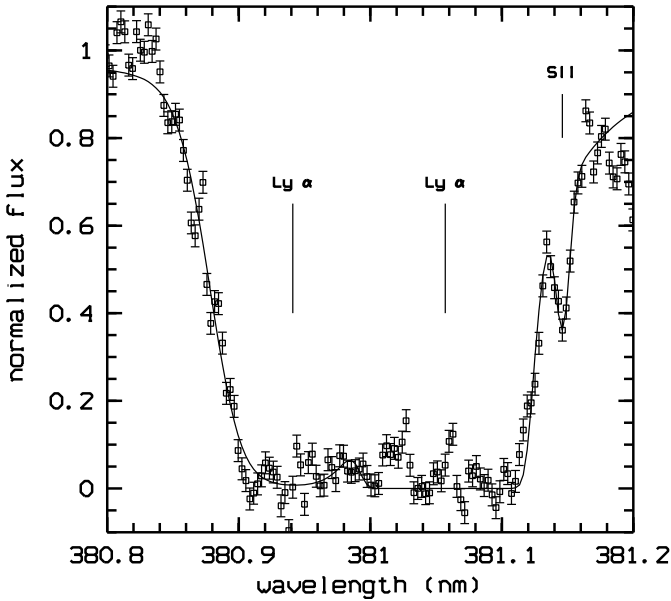,clip=t}
\caption{UVES spectrum of 0347-3819: the S{\sc ii} 94.7 nm
line of the absorption system at \zabs 3.02486, is clearly blended
with Ly$\alpha$ absorbers of lower redshift;
the continuous line is our best fit to the data.}
\label{q0347_2l}
\end{figure}

The line--of--sight towards  QSO 0347-3819 is characterized by the
presence of one damped Ly$\alpha$ system 
with two components very close to each other. The H{\sc i} column density
has been derived by Pettini et al (1994), while the
S{\sc ii} column density has been measured 
from the \ll 125.9 nm line by
Centuri\'on et al (1998),
Ledoux et al (1998) and Prochaska \& Wolfe (1999).
The measurements of Centuri\'on et al (1998) and Prochaska \& Wolfe (1999)
agree with each other, within errors, while the value
of Ledoux et al (1998) is lower by $\sim$ 0.3 dex.
We used the UVES spectrum
of this QSO to provide an external check on the accuracy
of our $f$ value, namely by comparing the column
density derived by the \ll 94.7 nm line with the 
values of Centuri\'on et al (1998) and Prochaska \& Wolfe (1999).
In Fig. \ref{q0347_2l} we show the observed spectrum
together with our best fit.
The S{\sc ii} \ll 94.7 nm line falls on the wing of a Ly$\alpha$
absorber of lower redshift
and this needs to be taken into account
when modeling the line. 
We fit the feature  successfully using two Ly$\alpha$
clouds and one S{\sc ii} \ll 94.7 nm line, as shown
in the figure.
The fitted parameters are log $N$(S{\sc ii})$=14.71\pm  0.04$,
$b=3.4\pm  1.0$. 
Another possible choice
is to fit only the red wing of the feature
by using one Ly$\alpha$ in this case we obtain
 log $N$(S{\sc ii}) 
$= 14.70\pm  0.04 $, the redshift is $z=3.024866$,
$b=3.1\pm 1.0$ \kms. 
Both the values
of the  column density found are  in perfect
concordance, within errors, with the value of
Centuri\'on et al (1998): log $N$(S{\sc ii}) $= 14.77\pm 0.08$
derived from CASPEC spectra using Voigt profile fitting, as we
do.
They also agree  with the value of Prochaska \& Wolfe (1999):
log $N$(S{\sc ii}) $= 14.731 \pm 0.012$ derived
from Keck HIRES spectra using the apparent optical depth method.
Our $b$ value is very small when compared
with $b=17.9 \pm 3.3$ \kms found by
Centuri\'on et al (1998). One should note however two facts:
in the first place Centuri\'on et al (1998) had a resolution
of 15.4 \kms , to be compared with 7.1  \kms of the present
data; in the second place, because of the lower resolution
Centuri\'on et al could not resolve the two components at
\zabs = 3.02466 and \zabs = 3.02486 but fitted
a single component with \zabs = 3.02476, i.e. the sum  of  the two.
In the UVES spectra the two components
are separated as may be appreciated
by inspecting several transitions, 
however for the \ll 94.7 nm  line the bluemost
component falls in the Ly$\alpha$ absorption and is not
detectable. Therefore our column density refers 
to the  \zabs = 3.02486 system only, while the value Centuri\'on et al
refers to the sum of both. It is thus not surprising that 
our column density is slightly {\it lower} than that 
of Centuri\'on et al. 
The agreement of our new measurement with those
of Centuri\'on et al (1998) and  Prochaska \& Wolfe (1999),
makes it unlikely that the abundance found 
by Ledoux et al (1998) is the correct one.  
 
To test the influence of the adopted $f$ on the derived
column density we performed fits using the two extreme values
given in the first row of Table 4.
For $f=0.00670$ we obtain log $N$(S{\sc ii}) $=14.57\pm  0.04$,
for $f=0.00360 $ we obtain log $N$(S{\sc ii}) $=15.11\pm  0.04$.
The fact that both values are in stark
disagreement with the column density
of Centuri\'on et al (1998) suggests that the $f$ value 
is more accurate than the formal $1 \sigma $ errors derived 
from the $\chi^2$ analysis.

\section{Application to the study
of BR J0307-4945}

\begin{figure}
\psfig{figure=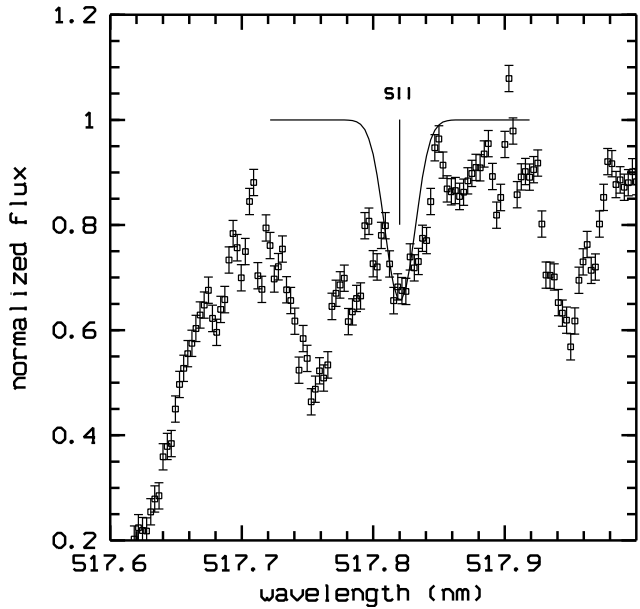,clip=t}
\caption{UVES spectrum of BR J0307-4945: the S{\sc ii} \ll 94.7 nm
line of the absorption system at \zabs 4.4680;
the continuous line is our best fit to the data.}
\label{q0305_2}
\end{figure}

\begin{figure}
\psfig{figure=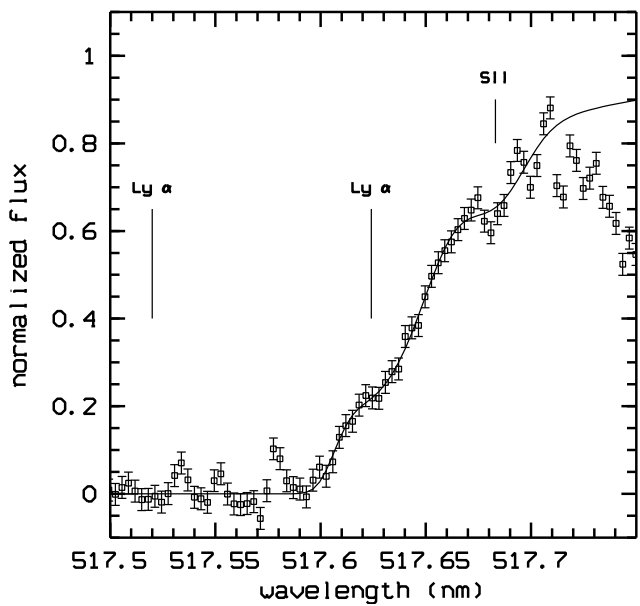,clip=t}
\caption{UVES spectrum of BR J0307-4945: the S{\sc ii} 94.7 nm
line of the absorption system at \zabs 4.4666, is blended
with lower redshift Ly$\alpha$ absorbers;
the continuous line is the our best fit to the data,
we modeled the Ly$\alpha$ absorption assuming two components.}
\label{q0305_1}
\end{figure}

This high redshift QSO ($z_{em}=4.74$) shows in its spectrum
the DLA at highest redshift so--far known: \zabs =4.466.
This very complex system, for which no less than 14 separate
components are required to describe the low--ions, is studied
in detail by Dessauges-Zavadsky et al (2001).
We identify the
S{\sc ii} \ll  94.7 nm line at \zabs = 4.4666 (component \# 8)
and at \zabs = 4.4680 (component \# 13), which are the main
components of the system.

The S{\sc ii} \ll  94.7 nm line at \zabs = 4.4680
is relatively blend-free, although there is clearly some
extra-absorption on the red wing.
The fitted parameters
are log $N$ (S{\sc ii}) $=14.72\pm  0.07$,
$b=7.7\pm  2.0$ and redshift $z=4.4681$, the fit is shown in
Fig. \ref{q0305_2}.
The sulphur abundance for this system, which has log $N$(H{\sc i}) $=20.56$
(Dessauges-Zavadsky et al 2001),
is therefore [S/H]$= -1.04$\footnote{we use here the usual
notation: $\mathrm [X/Y]= \log (X/Y)-\log (X/Y)_\odot$.}, 
which is consistent
with the upper limit given by Dessauges-Zavadsky et al ([S/H]$<-0.41$). 
By combining our measure of S column density, with the
column densities of Dessauges-Zavadsky et al for O, Si and Fe we obtain:
[S/O]$= +0.68$, 
[S/Si]$=+0.94$ and [S/Fe]$= +1.33$.
The above abundance ratios are  very peculiar.
Non solar S to $\alpha$ element ratios are
not theoretically excluded, although observationally this
would be the first occurence of such a case.   
Most likely  the line could be 
contaminated by a Ly$\alpha$, in which case our measure should
be considered as an upper limit to the S abundance.

The line of the \zabs = 4.4666 system
is blended with Ly$\alpha$ absorbers of lower redshift.
We obtain a reasonable fit   corresponding
to log $N$ (S{\sc ii}) $=14.47 \pm 0.26$,
$b=10.3\pm  3.9$ and redshift
$z=4.4667$. The result depends strongly on the modeling of
the Ly$\alpha$ absorption, therefore, although the fitted
redshift is in good agreement with that of component \# 8 of
Dessauges-Zavadsky et al (2001), we consider this
as a tentative identification.

\section{Conclusions}

By requiring that the S{\sc ii} \ll 94.7 nm  
line and the S{\sc ii} \ll 125 nm  triplet
provide the same S{\sc ii} column density 
in the line of sight towards HD 93521
we have been able to deduce an 
``astrophysical'' $f$ value for the S{\sc ii} \ll 94.7 nm line.
The error, of the order of 30\%, 
is  essentially associated to the noise present in the spectra.
As an external check of the accuracy of our $f$ value
we used the S{\sc ii}  \ll 94.7 nm at \zabs =3.02486 towards
QSO 0347-3819 to determine
the S{\sc ii} column density and were able to confirm, within errors, 
the 
value found from the \ll 125.9 nm  line by 
Centuri\'on et al (1998). This result gives us further confidence
in  our $f$ value
and suggests that the accuracy is better
than these formal error estimates. 
We warn the readers that the value log $gf = -0.230$ listed for the 
S{\sc ii} \ll 94.7 nm 
in table 3 of Barnstedt et al (2000) is a guessed value which
comes from the Kurucz (1993) 
database.  This value
is almost a factor of 30 
larger than our ``astrophysical'' value (log  $gf =-1.70$)
and should not be used in sulphur abundance analysis.

As an application we derived
the S{\sc ii} column density
in the absorption system at  \zabs = 4.4680
towards BR J0307-4945. In this system the S column density
cannot be derived from the  \ll 125 nm triplet because of blending
with Ly$\alpha$ clouds. The high [S/O] ratio implied by our
measurement has no straightforward explanation, which suggests
that the line could contaminated.

\section*{Acknowledgements}
We are grateful to Dr. Kappelmann for providing the ORFEUS
data and to Dr. Dessauges-Zavadsky and Dr. D'Odorico for
sharing with us the reduced UVES data.

\end{document}